\documentclass[aps]{revtex4}
\usepackage{graphicx}
\usepackage{dcolumn}
\usepackage{bm}
\usepackage{color}
\usepackage{eso-pic}
\usepackage{amsmath}
\usepackage{xcolor}
\usepackage{xspace}
\newcommand{\KTO}{$\mathrm{KTaO}_3$\xspace}
\newcommand{\STO}{$\mathrm{SrTiO}_3$\xspace}
\newcommand{\LAO}{$\mathrm{LaAlO}_3$\xspace}
 \usepackage{soul}

\def\e{\varepsilon}

\def\o{\omega}

\def\s{\sigma}

\def\lb{\label}
\def\pref#1{(\ref{#1})}
\begin{document}

\title{Superfluid stiffness of a \KTO -based two-dimensional electron gas}

\author{S. Mallik$^{1*}$, G. M\'enard$^{2*}$, G. Sa\"iz$^{2*}$, H. Witt$^{1,2}$, J. Lesueur$^{2}$, A. Gloter$^{3}$, L. Benfatto$^{4}$, M. Bibes$^{1}$, N. Bergeal$^{2}$ }

\affiliation{$^1$Unit\'e Mixte de Physique, CNRS, Thales, Universit\'e Paris-Saclay, 1 Avenue Augustin Fresnel, 91767 Palaiseau, France.}
\affiliation{$^2$ Laboratoire de Physique et d'Etude des Mat\'eriaux, ESPCI Paris, PSL University, CNRS, Sorbonne Universit\'e, Paris, France.}
\affiliation{$^3$Laboratoire de Physique des Solides, Universit\'e Paris-Saclay, CNRS UMR 8502, 91405 Orsay, France.}
\affiliation{$^4$Department of Physics and ISC-CNR, Sapienza University of Rome, P.le A. Moro 5, 00185 Rome, Italy.}
\affiliation{$*$S. Mallik, G. M\'enard and G. Sa\"iz contributed equally to this work.}

\maketitle
\large 
\textbf{ After almost twenty years of intense work on the celebrated \LAO/\STO system, the recent discovery of a superconducting two-dimensional electron gases (2-DEG) in (111)-oriented \KTO-based heterostructures injects new momentum to the field of oxides interfaces. However, while both interfaces share common properties, experiments also suggest important differences between the two systems. Here, we report gate tunable superconductivity in 2-DEGs generated at the surface of a (111)-oriented \KTO crystal by the simple sputtering of a thin Al layer. We use microwave transport to show that (111)-\KTO  2-DEGs exhibit a node-less superconducting order parameter with a gap value significantly larger than expected within a simple BCS weak-coupling limit model. Consistent with the two-dimensional nature of superconductivity, we evidence a well-defined Berezinsky-Kosterlitz-Thouless type of transition, which was not reported on \STO-based interfaces. Our finding offers innovative perspectives for fundamental science but also for device applications in a variety of fields such as spin-orbitronics and topological electronics.}\\

Potassium tantalate \KTO is a band insulator with a 3.6 eV gap that retains a cubic perovskite structure down to the lowest temperature \cite{fujii}. 
Like strontium titanate (\STO), it is a quantum paraelectric material on the verge of a ferroelectric instability that is characterized by a large permittivity at low temperature ($\epsilon_r\simeq$ 5000) \cite{fujii,fleury}. Both materials can be turned into a metal by electron doping, through oxygen vacancies for example. Because of their common properties, it was suggested that superconductivity should also occur in doped \KTO. However, while superconductivity was discovered more than half a century ago in bulk \STO \cite{schooley},  all the attempts to induce bulk superconductivity in \KTO have failed so far \cite{thompson}. Using ionic gating, Ueno \textit{et al.}  could generate a superconducting 2-DEG at the surface of (001)-\KTO albeit at very low temperature  ($T_c$ $\simeq$ 40 mK) \cite{ueno}. Later explorations of \KTO 2-DEGs did not evidence any superconductivity until the beginning of the year 2021 when two articles reported the discovery of superconducting 2-DEG formed at the interface between (111)-\KTO and insulating overlayers of \LAO or EuO \cite{liu,chen}. An empiric increase of $T_c$ with electron density was proposed with a maximum value of 2.2 K for a doping of $\approx$ 1.04 $\times$~$10^{14}$ $e^{-}\cdot\mbox{cm}^{-2}$ \cite{liu},  which is almost one order of magnitude higher than in the \LAO/\STO interface \cite{Reyren:2007p214}.  An electric field effect control of the $T_c$ was also demonstrated in a Hall bar device \cite{chen} and a dome-shaped superconducting phase diagram  similar to that of \STO-based interfaces was derived \cite{Caviglia:2008p116,biscaras2}. Following this discovery, the (110)-oriented \KTO interface was also found to be superconducting  with $T_c$ $\simeq$1 K \cite{chenPRL}.\\ 

\indent {In conventional superconductors, well described by the Bardeen-Cooper-Schrieffer (BCS) theory, the superconducting transition is controlled by the breaking of Cooper pairs as the temperature exceeds the energy scale set by the superconducting gap. However,  in two-dimensional superconductors the superfluid stiffness, i.e. the energy associated with the phase rigidity of the superconducting condensate, can be comparable to the pairing energy, allowing for a $T_c$ suppression driven by the loss of phase coherence.  In this case,  the transition is expected to belong to the Berezinsky-Kosterlitz-Thouless (BKT) universality class, where the transition is controlled by the unbiding of topological vortex-antivortex pairs \cite{bkt,bkt1,bkt2}.} Critical magnetic field measurements in (111)-\KTO 2-DEG, both in the perpendicular and in the parallel geometry, set an upper bound, $d$ $\approx$ 5 nm, on the extension of the 2-DEG in the substrate \cite{liu}. This is lower than the superconducting coherence length, $\xi$ $\approx$ 10-15 nm \cite{liu}, which confirms that the superconducting 2-DEG is in the 2D limit. In addition, the presence of disorder, which has been identified in this system \cite{liu,chen}, is also expected to lower the superfluid rigidity and reinforce the role of phase fluctuations. Even though the measurements of the current-voltage characteristics  in Ref.\  \cite{liu} could be compatible with indirect signatures of a BKT transition, a direct measurement of the superfluid stiffness is required to properly address this issue \cite{venditti_prb19}.\\
	
	Here, we show that a 2-DEG can be generated at the surface of a (111)-oriented \KTO crystal simply by sputtering a very thin Al layer.  The deposition of Al leads to the reduction of Ta ions as evidenced by X-ray photoelectron spectroscopy (XPS) and leads to the formation of an interfacial gate tunable superconducting 2-DEG.  We use resonant microwave transport to measure the complex conductivity of the  2-DEG and extract the temperature dependent superfluid stiffness.  The superconducting 2-DEG exhibits a  node-less order parameter with a  gap value significantly larger than expected within a simple BCS weak-coupling limit model and the superconducting transition follows the Berezinsky-Kosterlitz-Thouless model, which was not observed on \STO-based interfaces.

\begin{figure}[t]
\vskip 0.5cm
\includegraphics [width=9cm]{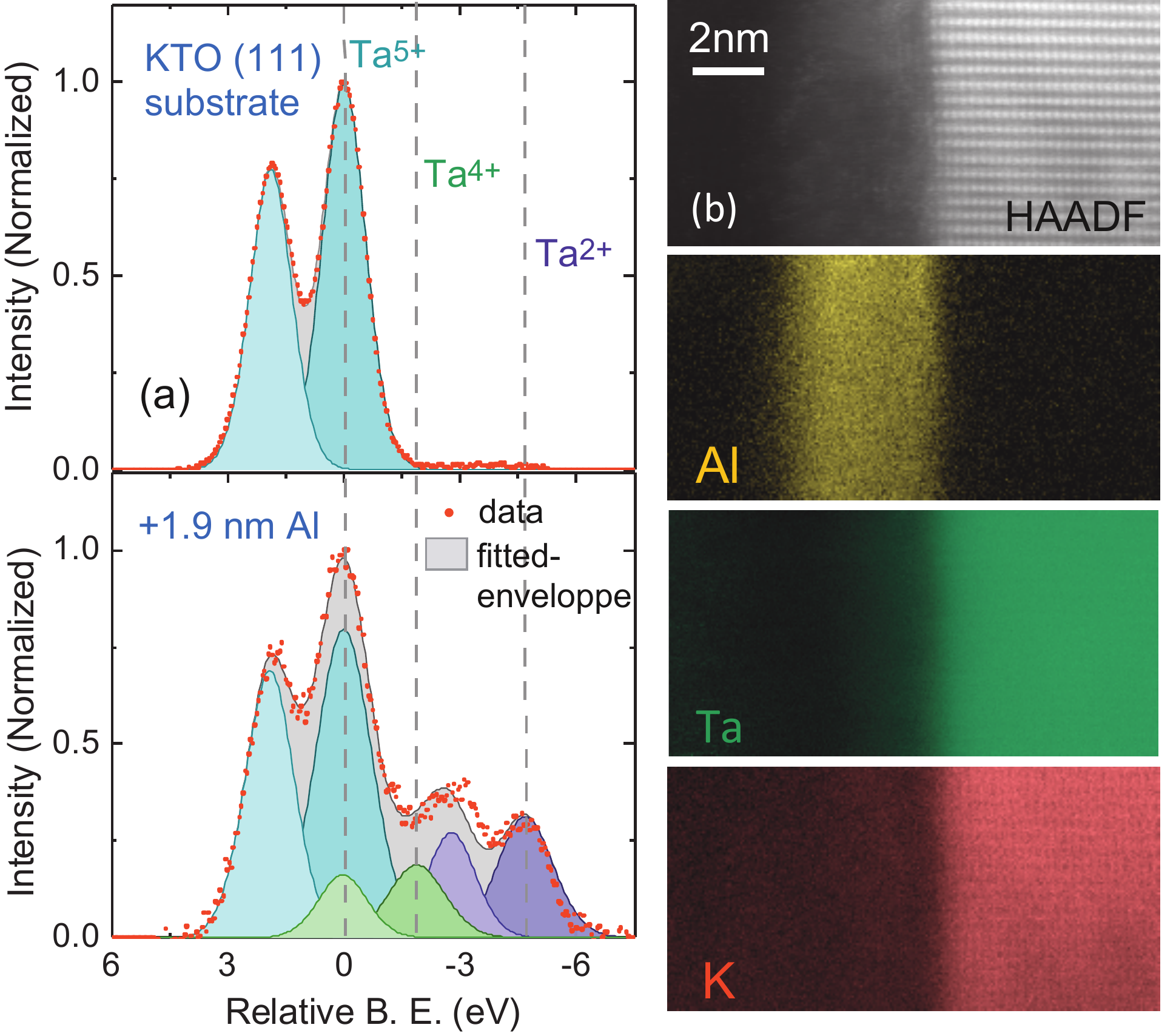}%
\vskip -0.5cm
\caption{\textbf{XPS and STEM characterization of the AlOx/\KTO samples}. (a) X-ray photoelectron spectra near the 4f core level binding energy of Ta for a \KTO substrate prior to deposition (top) and after deposition of 1.8 nm of Al (bottom). The fitted peaks for Ta$^{5+}$, Ta$^{4+}$ and Ta$^{2+}$ are shown in cyan, green and violet colors, respectively. The deeper and lighter shades of same colors represent the 4f$_{5/2}$ and 4f$_{2/2}$ valence states of the respective peaks. The data and sum fit envelope are shown in red circles and black line. (b) (top) HAADF scanning transmission electron microscopy  image at the cross-section of \KTO (111) and AlOx interface. The \KTO is observed along the [112] direction. (down) EELS maps (Al-\textit{L}$_{2,3}$ , Ta-\textit{O}$_{2,3}$, K-\textit{L}$_{2,3}$ edges) showing the presence of Al on top of the interface without any interdiffusion but with some limited diffusion of Ta and K inside the AlOx layer.}
\end{figure}

2-DEGs were generated by dc sputtering of a very thin Al layer on (111)-oriented \KTO  substrates at a temperature between 550 $^\circ$C and 600 $^\circ$C. The preparation process is detailed in the Methods section. Prior to deposition, we measured the in-situ X-ray photoelectron spectra (XPS) of the Ta 4f valence state (Fig. 1a top) of the \KTO substrate. The spectra shows the sole presence of Ta$^{5+}$ states (4f$_{5/2}$ and 4f$_{3/2}$) indicating the expected stoichiometry of the substrate. The Ta 4f core levels were then measured after growing 1.8-1.9 nm of Al and transferring the sample in vacuum to the XPS setup. The bottom graph in Fig. 1a shows the Ta 4f core level spectra with additional peaks corresponding to reduced states of Ta i.e., Ta$^{4+}$ and Ta$^{2+}$. The deeper and lighter shades of same-coloured peaks correspond to 4f$_{5/2}$ and 4f$_{3/2}$ splitted peaks. The reduction of Ta$^{5+}$ to Ta$^{4+}$ and Ta$^{2+}$ upon Al deposition indicates the formation of oxygen vacancies at the surface of \KTO, which in turn suggests the formation of a 2-DEG. We monitored the Al oxidation state by measuring the Al 2p core levels after exposure of the sample to the atmosphere, which evidenced a full oxidation of the Al layer into AlOx. Thus, as in the AlOx/\STO system, the 2-DEG formed through a redox process by which oxygens are transferred from the \KTO substrate to the Al overlayer \cite{luisPRM,vaz,rodel}.\\
\indent The structure of the AlOx/\KTO(111) interface has been imaged by scanning transmission electron microscopy (STEM). Fig. 1b depicts the high-angle annular dark field (HAADF) - STEM image in cross-section. The electron energy loss spectroscopy (EELS) indicates that a small amount of K and Ta diffuse into the AlOx layer.  In contrast, the Al signal decays very rapidly in \KTO indicating no Al diffusion into \KTO. Our fabrication method based on the sputtering of a thin Al film has already been successfully implemented to generate 2-DEGs on (001)-oriented \KTO substrate showing a fivefold enhancement of the Rashba spin-orbit coupling as compared to \STO \cite{luis}.
In the present work, four samples, labelled \textbf{A}, \textbf{B}, \textbf{C} and \textbf{D}  have  been investigated by transport measurement at low temperature in a dilution refrigerator (see Methods section for fabrication parameters).\\

\begin{figure}[t]
\vskip 0.5cm
\includegraphics [width=12cm]{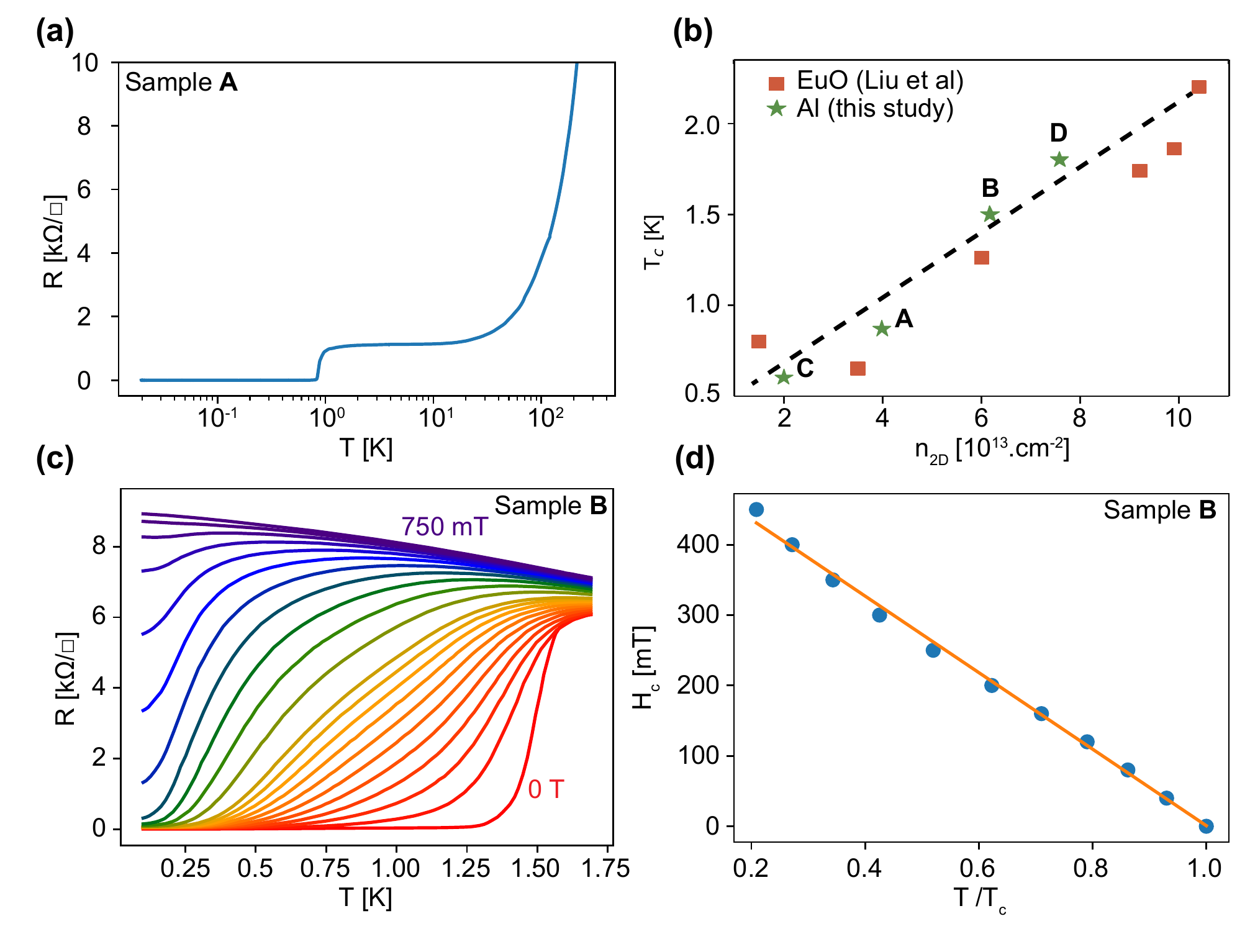}%
\vskip -0.5cm
\caption{\textbf{Magneto-transport characterization}. (a) Sheet resistance of sample \textbf{A} as a function of temperature (log scale) showing a superconducting transition at $T_c$ $\simeq$ 0.9 K. (b) Summary of the superconducting critical temperature as a function of the carrier density for all samples studied in this work compared with results from \cite{liu}. (c) Sheet resistivity of sample \textbf{B}  as a function of temperature for increasing magnetic fields between 0 and 750 mT. (d) Perpendicular critical magnetic field defined as the temperature at which the resistance recover 80 $\%$ of the normal state resistance measured at T = 1.65 K. The orange line correspond to a linear fit with Ginsburg Landau formula.}
\end{figure}

 Fig. 2a shows the resistance vs temperature curve of sample \textbf{A} on a wide temperature range revealing a superconducting transition at $T_c$ $\simeq$ 0.9K. In Fig. 2b, we plot the $T_c$ as a function of the 2D carrier density, $n_\mathrm{2D}$ for the different samples studied and compare their values with those extracted from reference \cite{liu}. Our results confirm the trend observed in the literature ($T_c$ increases with the carrier density) and demonstrates that our growth method, while being much easier to implement than the molecular beam epitaxy of a rare-earth element such as Eu or the pulsed laser deposition of a complex oxide, is able to produce good quality samples with similar $T_c$. The resistance vs temperature curves of sample \textbf{B}  measured for different values of a magnetic field applied perpendicularly to the sample plane are shown in Fig. 2c. The temperature dependence of the critical magnetic field is consistent with a Landau-Ginsburg model near $T_c$, $H_c(T)=\frac{\Phi_0}{2\pi\xi_\parallel^2(T)}$, taking into account an in-plane superconducting coherence length $\xi_\parallel=\xi_\parallel(0)(1-\frac{T}{T_c})^{-\frac{1}{2}}$. We found $\xi_\parallel(T=0)$ $\approx$ 24 nm, which is comparable with the value reported in reference \cite{liu}.\\ 
 
\indent Although \KTO is a quantum paraelectric material  like \STO, its  permittivity  is reduced by a factor five as compared to \STO making electric field effect less efficient in a back-gating configuration  \cite{fujii,fleury}. To overcome this difficulty, we prepared a AlOx/\KTO sample using a thinner substate (150 $\mu$m). After cooling the sample, the back-gate voltage was first swept to its maximum value $V_\mathrm{G}$ = 200 V while keeping the 2-DEG at the electrical ground. This forming procedure ensures the reversibility of the gate sweeps in further gating sequences \cite{biscaras3}. Fig. 3 shows the sheet resistance of sample \textbf{C} as a function of temperature for different values of the gate voltage between -40 V and 200 V.  Electrostatic gating induces both a  modulation of the normal-state resistance  and a variation of the superconducting critical temperature. For negative gate voltages, corresponding to a depleted quantum well, R vs T curves exhibit a quasi-reentrant behaviour : the resistance first decreases and then upturns upon further cooling  \cite{jaeger,orr}. This is characteristic of disordered superconducting thin films in which superconductivity only exists locally, forming a network of  isolated islands  surrounded by an insulating medium that precludes percolation. While the decrease of resistance marks the emergence of superconductivity  inside the islands, the upturn of resistance at low temperature results from the opening of a gap in the excitation spectrum, which prevents the flow of quasiparticles across islands. Hence, the resistance does not reach zero, indicating that the superconducting order does not extend at long range. As carriers are added  upon increasing the gate voltage, the resistance curves flatten at low temperature and the 2-DEG eventually reaches a true zero resistive state ($V_\mathrm{G}$ $>$ -25V).  A long range superconducting order is established through Josephson coupling between the islands.  Further doping makes the network of islands denser and increases the coupling between islands resulting in a ``homogeneous-like" superconducting 2-DEG at high-doping.   The resulting superconducting phase diagram is shown in Fig. 3b, where the resistance is plotted in color scale as a function of temperature and electron density extracted by combining Hall effect and gate capacitance measurement \cite{singhPRB,biscaras2}. It ends by a quantum critical point in the depleted region. In this experiment, the carrier density was tuned from $n_\mathrm{2D}\simeq 1\times$~$10^{13}$  $e^{-}\cdot\mbox{cm}^{-2}$ to $n_\mathrm{2D}\simeq 2.2\times$~$10^{13}$  $e^{-}\cdot\mbox{cm}^{-2}$, which is not sufficient to explain the modulation of the normal resistance by more than one order of magnitude. This indicates that the gate voltage not only controls the carrier density, but also modifies deeply the electronic properties of the 2-DEG, in particular the electronic mobility, in agreement with previous reports \cite{chen}.

  \begin{figure}[t]
\vskip 0.5cm
\includegraphics [width=10cm]{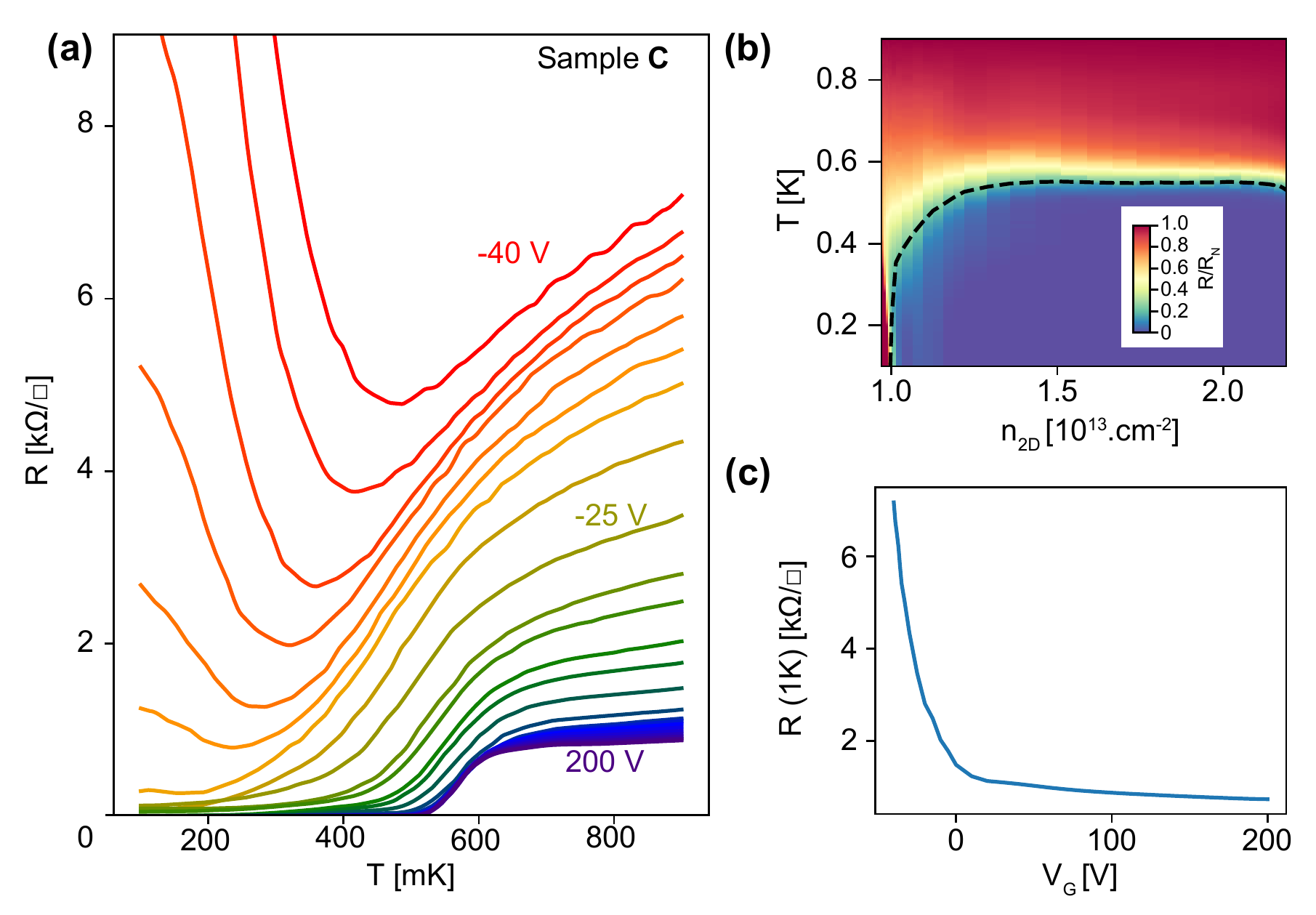}%
\vskip -0.5cm
\caption{\textbf{Electric field effect and superconducting phase diagram}. (a) Temperature dependence of the sheet resistance of sample \textbf{C}  for difference values of the back-gate voltage  in the range -40 to 200~V. (b) Normalized sheet resistance in color scale as a function of the carrier density of extracted from Hall measurements and temperature. The dashed line indicates the critical temperature defined by a 80$\%$ drop of the resistance with respect to the normal resistance $R_N$ at T=0.9K.  (c) Sheet resistance at 0.9~K as a function of the back-gate voltage.}
\end{figure}

We further investigated the  superconducting  \KTO 2-DEG by measuring its superfluid stiffness $J_{s}$, which is the energy scale associated with the phase rigidity of the superconducting condensate. $J_{s}$ is related to the imaginary part  of the complex conductivity $\sigma(\omega)=\sigma_1(\omega)-i\sigma_2(\omega)$ of the superconductor that accounts for the transport of Cooper pairs at finite frequency $\omega$. This is a direct probe of the superconducting order parameter that provides important information on the nature of the superconducting state.  In the low frequency limit, $\hbar\omega\ll\Delta$, a superconductor displays an inductive response to an ac electrical current and $\sigma_2(\omega)$=$\frac{1}{L_k\omega}$, where $L_k$ is the kinetic inductance of the superconductor that diverges at $T_c$ \cite{tinkham}. The superfluid stiffness is then directly related to $L_k$
\begin{eqnarray} 
J_s(T)=\frac{\hbar^2}{4e^2L_k(T)}
\end{eqnarray}  
where $\hbar$ is the reduced Planck constant and $e$ is the electron charge.\\

We used resonant microwave transport to extract $L_k$ below $T_c$ and determine the superfluid stiffness of sample \textbf{D} as a function of temperature. The method, which was successfully applied to superconducting \STO-based interfaces, is illustrated in Fig. 4a and described in details in references \cite{singh, singhNM}. In short, the \KTO sample is embedded into a parallel RLC resonant electrical circuit made with Surface Mount microwave Devices (SMD). The capacitance of the circuit is  dominated by the \KTO substrate contribution ($C_\mathrm{KTO}$) due to its large intrinsic permittivity. The total inductance of the circuit, $L_\mathrm{tot}(T)=\frac{L_1L_\mathrm{k}(T)}{L_1+L_\mathrm{k}(T)}$, includes the contribution of a SMD inductor ($L_{1}$) and the contribution of the kinetic inductance $L_k$ of the superconducting 2-DEG below $T_c$. Finally, a SMD resistor $R_1$ imposes that the dissipative part of the circuit impedance remains close to 50 $\Omega$ in the entire temperature range, ensuring a good impedance matching with the microwave circuitry. The circuit resonates at the frequency $\omega_{0}=\frac{1}{\sqrt{L_\mathrm{tot}C_\mathrm{KTO}}}$, which is accessed by measuring the reflection coefficient  of the sample circuit $\Gamma(\omega)=\frac{A^\mathrm{in}}{A^\mathrm{out}}=\frac{Z(\omega)-Z_0}{Z(\omega)+Z_0}$. The resonance manifests itself as a peak in the real part of  the circuit impedance, $Z(\omega)$, accompanied by a $\pi$ phase shift \cite{singh}. The height and the width of the peak are controlled by the dissipative part of the circuit impedance.  In the superconducting state, the 2-DEG conductance acquires a kinetic inductance $L_\mathrm{k}$ that generates a shift of $\omega_0$ towards high frequencies with respect to the normal state (Fig. 4b).  The temperature dependent superfluid stiffness $J^\mathrm{exp}$, extracted from the resonance shift and Eq. (1) is presented in Fig. 4c (blue circles). \\

The flattening of the $J^\mathrm{exp}$ curve below 1 K indicates a fully gapped behavior, i.e. an absence of nodes in the order parameter.  
  The purpled dashed line ($J^\mathrm{BCS}$) shows an attempt to fit the experimental curve with a standard BCS expression $J^\mathrm{BCS}_s(T)/J_s(0)=(\Delta(T)/\Delta(0))\tanh (\Delta(T)/k_\mathrm{B}T)$ \cite{tinkham}, where $\Delta(T)$ is the superconducting gap obtained numerically by self-consistent solution of the BCS equation, so that it vanishes at the mean-field temperature $T_c^0$ (i.e. the temperature at which Cooper pairs form). Since $J_s(0)$ is fixed by the experimental value at the lowest temperature, the only free parameter is then the ratio $\Delta(T=0)/k_\mathrm{B}T^0_c$, that determines the curvature of the $J_s(T)$ curve.  As one can see, even using a relatively strong-coupling value $\Delta(0)/k_\mathrm{B}T^0_c$ = 2.3, from the fit of the low-temperature curve one obtains $T_c^0$ $\simeq$ 2.2 K, that is larger than the experimental $T_c$. To fit the data in the whole temperature range with the BCS expression only, one would then need an unreasonably large value ($\Delta(0)/k_\mathrm{B}T^0_c$ $\simeq$ 6). Here we follow a different approach and interpret the rapid drop of $J_s(T)$ below the BCS fit as a BKT signature, as we will discuss below. This interpretation is supported by a second striking observation, that holds regardless of any specific consideration about its temperature dependence : the $T=0$ value of the stiffness $J_s(T=0)$ $\simeq$ 7.3 K is of the same order as $T_c$ $\simeq$ 2.2 K. It is worth noting that in conventional superconductors, where the superfluid density $n_s(T=0)$ is close to the carrier density $n_\mathrm{2D}$, the stiffness at zero temperature is of the order of the Fermi energy, and then several orders of magnitude larger than $T_c^0$. A strong reduction of $J_s(0)$ is instead observed in 2D-superconductors, where disorder strongly reduces $n_s$ with respect to $n_\mathrm{2D}$ already at $T=0$ \cite{epstein_prl81,epstein_prb83,fiory_prb83,lemberger_prl00,armitage_prb07,armitage_prb11,kamlapure_apl10,mondal_bkt_prl11,yazdani_prl13,yong_prb13,ganguly_prb15}. In the dirty limit, in which the elastic scattering rate $1/\tau$ is much larger than the superconducting gap, only a fraction of carriers , $n_s/n_\mathrm{2D} \simeq 2\Delta(0)/(\hbar/\tau$), forms the superconducting condensate. In a single-band picture, an estimate of the superfluid stiffness is obtained from $\Delta(0)$ and the normal resistance $R_n$,  $J_s\simeq \frac{\pi\hbar\Delta(0)}{4e^2R_n}$. Using the previously estimated value of $\Delta(0)$ $\simeq$ 5 K and $R_n$ $\simeq$ 1300 $\Omega$, we obtain $J_s$ $\simeq$ 11.8 K close to the measured value ($J^\mathrm{exp}_s(T=0)$ $\simeq$ 7.3 K), which is consistent with the dirty limit.

  The superfluid density of the 2-DEG can be directly deduced from the stiffness through the formula $n_\mathrm{s}=\frac{4m}{\hbar^2}J_\mathrm{s}$, where $m$ is the effective mass of superconducting electrons. In the case of (111)-\KTO 2-DEGs, the conduction band is derived from the bulk J = 3/2 states with a Fermi Surface formed by  an hexagonal contour inside a sixfold symmetric star-shaped contour  \cite{bruno}.  Considering  an average effective mass $m$ $\simeq$ 0.5$m_0$, the corresponding superfluid density $n_s$ extracted from $J_s^\mathrm{exp}$  is $n_s$ $\simeq$ 1.8 $\times$ 10$^{12}$~e$^{-}$.cm$^{-2}$, which is about $2.5\%$ of the total carrier density ($n_\mathrm{2D}$ = 7.5 $\times$ 10$^{13}$~e$^{-}$.cm$^{-2}$  for sample \textbf{D}). This very low ratio is comparable with previous findings in \LAO/\STO  interfaces \cite{singh,caviglia_cm18,bert_prb12}. Although such reduced superfluid density is consistent with the dirty limit, \KTO-(111) 2-DEG is a multiband system \cite{bruno}, in which superconductivity may involve only specific bands as also suggested in \STO \cite{singh}. \\ 
  
 The reduced dimensionality and the suppression of the energy scale associated with the stiffness represent the prerequisite to observe BKT \cite{bkt,bkt1,bkt2} physics, since it makes the BKT temperature scale $T_\mathrm{BKT}$ associated with the unbinding of vortex-antivortex pairs far enough from $T_c^0$  \cite{benfatto_review14}. The most famous hallmark of the BKT transitions is the discontinuous jump to zero of $J_s$ at $T_\mathrm{BKT}<T_c$  with an universal ratio
$J_s(T_\mathrm{BKT})/T_\mathrm{BKT} = 2/\pi$ \cite{nelson}. Such a prediction, theoretically based on the study of the 2D $XY$ 
model \cite{bkt,bkt1,bkt2},  has been successfully confirmed in superfluid He films \cite{helium4}.  In practice, the experimental observation of the BKT transition in real superconductors is more subtle. 
Indeed, in thin films the suppression of $n_s$ (and then $J_s$) with disorder comes along with an increasing {\em inhomogeneity} of the SC background, that is predicted to smear out the discontinuous superfluid-density jump \cite{benfatto_prb09,mondal_bkt_prl11,maccari_prb17} into a rapid downturn, as observed experimentally via the direct 
measurement of the inverse penetration depth  
\cite{lemberger_prl00,armitage_prb07,armitage_prb11,kamlapure_apl10,
mondal_bkt_prl11,yazdani_prl13,yong_prb13,ganguly_prb15} or indirectly via the measurement of the exponent of the non-linear $IV$ characteristics near $T_c$ 
\cite{epstein_prl81,epstein_prb83,fiory_prb83, venditti_prb19}. In the case of \STO-based interfaces, the direct 
measurement of $J_s$ is rather challenging, and the few experimental reports available so far
do not evidence a BKT jump \cite{bert_prb12,singh,caviglia_cm18}.\\

\begin{figure}[t]
\vskip 0.5cm
\includegraphics [width=14cm]{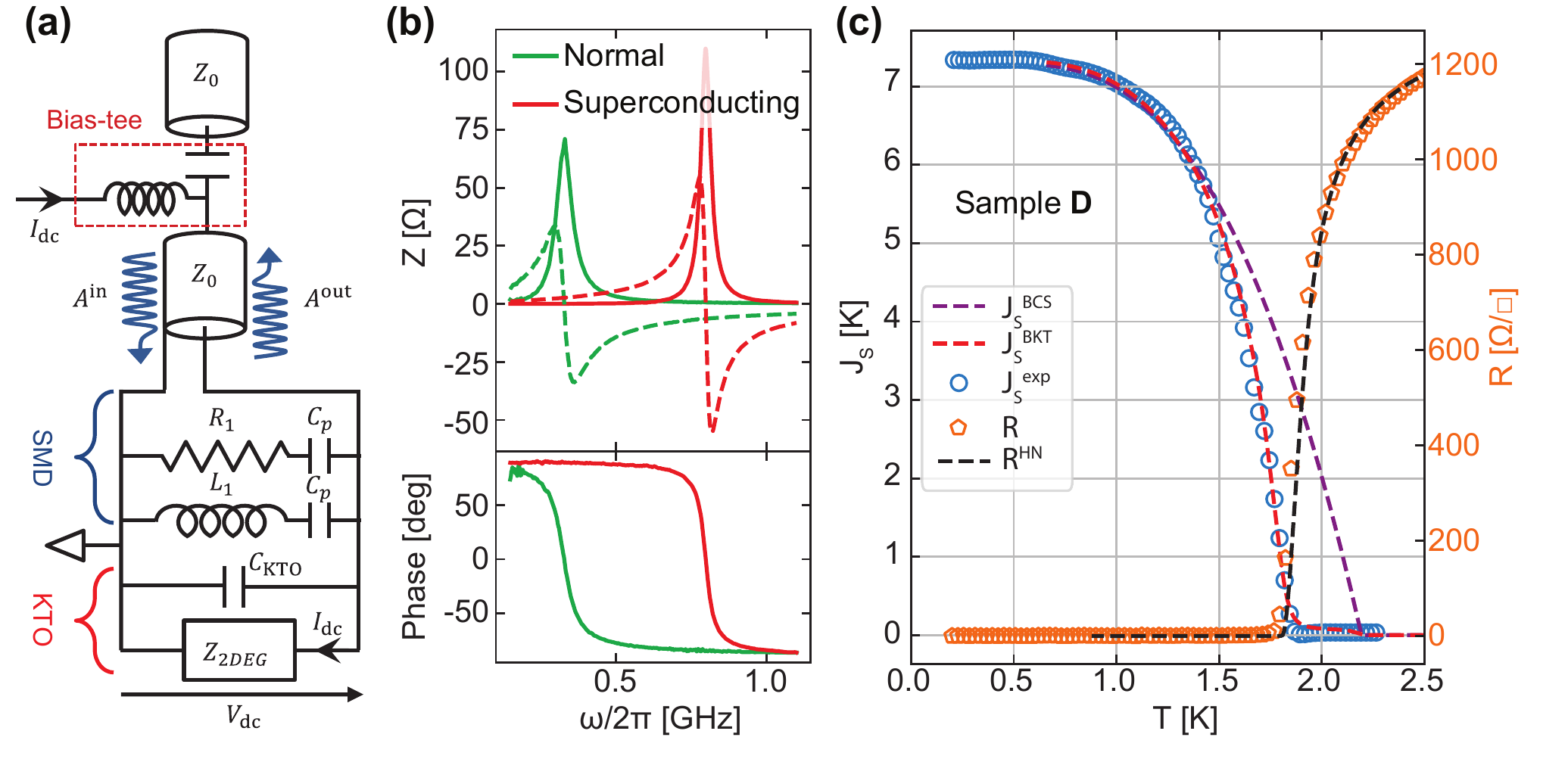}%
\vskip -0.5cm
\caption{\textbf{Microwave measurement of the \KTO 2-DEG}. (a) Schematics of the measurement set up adapted from \cite{singh}. The contribution of the sample is represented by the 2-DEG impedance, $Z_{2DEG}$,  and the parallel capacitive contribution of the \KTO substrate, $C_\mathrm{KTO}$. SMD components, resistor $R_1$=75 $\Omega$, inductor $L_1$=6.5nH,  are placed in parallel  to define a resonating circuit. Large SMD capacitors $C_p$=2$\mu_F$ block the dc signals in $L_1$ and $R_1$ without influencing the signals at microwave frequencies. The reflection coefficient at the circuit sample $\Gamma(\omega)$ = $\frac{A^{out}}{A^{in}}$ is extracted from the measurement as described in \cite{singh}. 
A bias-tee is used  to separate the dc current and the microwave one. (b) Top : Real (full lines) and imaginary (dashed lines) parts of the 2-DEG impedance as a function of frequency in the normal state at T = 2.5 K and in the superconducting state at T = 0.2 K after a calibration procedure \cite{singh}. Bottom : corresponding phases of the 2-DEG impedance in the normal and superconducting state. The resonance frequency of the sample circuit can be clearly identified, for instance as the maximum of the peak in the real part of the impedance or as a $\pi$ phase shift in its phase.
(c) Superfluid stiffness $J_{s}^\mathrm{exp}$  extracted from the resonance frequency and Eq. (1) as a function of the temperature. The dashed purple line shows and attempt to fit the experimental data within a standard BCS model ($J_{s}^\mathrm{BCS}$), which provides a mean field $T_c^0$ =2.3 K. A better  agreement is obtained by using a BKT model ($J_{s}^\mathrm{BKT}$). On the right axis, the figure also shows the sheet resistance curve ($R$) fitted with the Halperin and Nelson formula ($R_\mathrm{HN}$). }
\end{figure}

 Within the BKT approach, the effect of vortex-like topological excitations provides an additional suppression of $J_s$ with respect to the BCS dependence discussed above, driven only by quasiparticle excitations. To provide a fit of $J_s^\mathrm{exp}$,  we then solved numerically the renormalization-group (RG) equations of the BKT theory for the superfluid stiffness and vortex fugacity. As input parameters of the RG equations we used the BCS temperature dependence of the stiffness, and we additionally included the finite-frequency effects \cite{ambegaokar_prb80,HN, benfatto_review14}. Indeed, even though the resonance frequency (about $0.5$ GHz) is still small  as compared to the optical gap  ($2\Delta$ $\sim$ 10 K $\sim$ 200 GHz) it can nonetheless lead to non-negligible effects, in particular a rounding of the jump and a suppression of the stiffness at a temperature slightly larger than the one where the dc resistivity vanishes \cite{armitage_prb07,armitage_prb11,ganguly_prb15}, as indeed observed in our case. The resistivity itself is consistently fittted with the interpolating Halperin-Nelson formula\cite{HN}, which accounts for BKT-like fluctuations between $T_{BKT}$ and $T_c^0$, and for standard Gaussian fluctuations above $T_c^0$. Finally, to account for spatial inhomogeneities, we introduce a gaussian distribution of local $T_c$ and $J_s$ with variance $\sigma_G$ centered around $T_c^0$ and $J_s^\mathrm{exp}(0)$. As seen in Fig. 4c, the result of the fitting procedure (dashed red line) is in very good agreement with experimental data both for the superfluid stiffness and the resistance considering a very small inhomogeneity, $\sigma_G$=0.02. Details on the fitting procedure are given in the Method Section.

Although \KTO and \STO have many common properties, the superconducting phases of their interfacial 2-DEG exhibit noticeable differences. Whereas a pure BCS weak-coupling limit with $\Delta(0)/k_BT_c$ $\simeq$ 1.76 provides a very good description of superconductivity in \STO-based interfaces \cite{singh, singhNM}, we found a stronger value of the coupling for \KTO ($\Delta(0)/k_BT_c$ $\simeq$ 2.3). Such important difference, which must be traced back to the pairing mechanism, is a strong constraint on the possible origin of superconductivity in these two materials. In addition, BKT physics was not observable in \STO for which a simple BCS model without phase fluctuations was sufficient to fit the $J_s(T)$ curves with a very good accuracy \cite{singhNM}. This may suggest a more bosonic-like superconductivity in  \KTO-based interfaces (in the highly doped regime), as evidenced by the large separation between the pairing scale, set by $\Delta$, and the phase-coherence scale, set by the small value of the superfluid stiffness. Recent measurements of the in-plane critical field in \KTO-based interfaces suggested that the order parameter could be a mixture of s-wave and p-wave pairing components induced by strong spin-orbit coupling \cite{zhang}. While we can not rule out this possibility,  the saturation of the $J_s(T)$ curve below $T_c/2$ seen in Fig. 4b suggest a dominance of the fully gapped s-wave component. Further experiments, including tunneling spectroscopy are therefore necessary to understand the nature of superconductivity in \KTO-based interfaces.\\

\textbf{\large{Methods}}\\
\textbf{Sample fabrication}.
Prior to deposition, \KTO (111) substrates from MTI corporation were annealed at 600 $^\circ$C for 1 hour in vacuum. Then, the thin Al layer was deposited in a dc magnetron sputtering system (PLASSYS MP450S) under a base pressure of the vacuum chamber  lower than 5$\times$10$^{-8}$ mbar. During Al deposition, the Ar partial pressure and the dc power were kept fixed at 5$\times$10$^{-4}$  mbar and 10 W, respectively. The deposition rate for Al was  0.66 $\AA$/s. The table below summarizes the deposition parameters for the different samples.\\

\begin{tabular}{|c|c|c|c|}
  \hline
  Samples & Deposition temp.  & Al thickness & \KTO thickness \\
  \hline
  \textbf{A} & 600 $^\circ$C & 1.8 nm & 0.5 mm \\
  \textbf{B} & 500 $^\circ$C & 1.8 nm & 0.5 mm \\
   \textbf{C} & 600 $^\circ$C & 1.8 nm & 0.15 mm \\
    \textbf{D} & 500 $^\circ$C 1$^\mathrm{st}$ step
     RT 2$^\mathrm{nd}$ step  & 0.8 + 1.1 nm & 0.5 mm \\
   
  \hline
\end{tabular}\\

\textbf{XPS analysis}.
X-ray photoelectron spectroscopy (XPS) was measured using a non-monochromatized Mg K$_\alpha$ source ($h\nu$ = 1253.6 eV) in an Omicron NanoTechnology GmbH system with base pressure of 5$\times$10$^{-10}$ mbar. The operating current and voltage of the source was 20 mA and 15 kV, respectively. Spectral analysis to determine different valence states of Ta were carried out using the CasaXPS software. Adventitious carbon was used as a charge reference to obtain the Ta 4f$_{5/2}$ peak position for the fitting. The energy difference and the ratio of the area between 4f$_{5/2}$ and 4f$_{3/2}$ peaks for all the Ta valence states were constrained according to the previously reported values.\\

\textbf{STEM characterization}. STEM-HAADF and STEM-EELS measurements have been done at 100keV using a Cs corrected Nion STEM microscope and a Gatan modified EELS spectrometer equipped with a MerlinEM detector.\\ 

\textbf{Theoretical analysis of $J_s(T)$}. 
In order to account for vortex excitations we solved the BKT RG equations \cite{bkt2,nelson,benfatto_review14} for the vortex fugacity $g=2\pi e^{-\mu/(k_BT)}$, with $\mu$ the vortex-core energy,  and the rescaled stiffness $K\equiv \pi J_s/k_BT$:
\begin{eqnarray}
\label{eqk} 
\frac{dK}{d\ell}&=&-K^2g^2,\\ 
\label{eqg} 
\frac{dg}{d\ell}&=&(2-K)g,
\end{eqnarray}
where $\ell=\ln (a/\xi_0)$ is the RG-scaled lattice spacing with respect to the coherence length $\xi_0$, that controls the vortex sizes and appears as
a short-scale cut-off for the theory. The initial values at $\ell=0$ are set by the BCS fitting $J^\mathrm{BCS}(T)$ of $J_s^\mathrm{exp}$, and the renormalized stiffness is given by the large-scale behavior, $J_s=(k_B T/\pi)K(\ell\rightarrow \infty)$. The ratio $\mu/J_s=0.87$, similar to the one found in other conventional superconductors \cite{kamlapure_apl10,
mondal_bkt_prl11,yong_prb13,ganguly_prb15}, is used as a free (temperature-independent) parameter, which controls the strength of stiffness renormalization due to bound vortices below $T_\mathrm{BKT}$ \cite{benfatto_review14}. To account for finite-frequency effects we further include dynamical screening of vortices \cite{ambegaokar_prb80,HN}  via an effective
frequency-dependence dielectric function $\e(\o)$ which enters in the complex conductivity of the film as $\s(\o)=-\frac{4J^\mathrm{BCS} e^2}{i\o \hbar^2\e(\o)}$. At zero frequency $\e(\omega)$ is real and $\e_1(0)=K(0)/K(\ell\rightarrow\infty)=J^\mathrm{BCS}/J_s$ so one recovers the usual static result. At finite frequency $\e(\o)$ develops an imaginary part due to the vortex motion, that can be expressed in first approximation \cite{ambegaokar_prb80} as 
$\e_2\simeq ({r_\o}/{\xi})^2$, where $\xi$ is the vortex
correlation length and $r_\o$ is a finite length scale set in by the finite frequency of the probe, i.e.
$r_\omega=\sqrt{\frac{14 D_v}{\omega}}$,  with $D_v$  the vortex diffusion constant of the vortices. The main effect of $\e_2$ is to induce a small tail above $T_\mathrm{BKT}$ for the finite-frequency stiffness, as given by $J_s=\hbar^2 \omega\sigma_2(\omega)/(4e^2)$, as we indeed observe in the experiments. Here we follow the same procedure outlined in Ref.\ \cite{ganguly_prb15} to compute $\e(\omega)$, and in full analogy with this previous work we find a very small vortex diffusion constant $D_v$ $\sim$ 10$^{10}$ nm$^2$/s.  The correlation length $\xi(T)$ also enters the temperature dependence of the resistivity above $T_\mathrm{BKT}$, that follows the usual scaling $R/R_N={1}/{\xi^2(T)}$. To interpolate between the BKT and Gaussian regime of fluctuations we use the  well-known Halperin-Nelson expression \cite{HN,benfatto_prb09,benfatto_review14} $\xi_{HN}(T)=\frac{2}{A}\sinh \left(\frac{b}{\sqrt{t}}\right)$ where $t=(T-T_\mathrm{BKT})/T_\mathrm{BKT}$, and we set $A=2.5$ and $b=0.27$, consistent with the theoretical estimate of $b$ $\simeq$ 0.2 that we obtain from the value of $\mu$ \cite{benfatto_review14,mondal_bkt_prl11,ganguly_prb15}. Finally, to account for the possible inhomogeneity of the sample we consider the extension of the previous method to the case where the overall complex conductivity of the sample is computed 
in the self-consistent effective-medium approximation \cite{kirkpatrick}  as solution of the following equation:
\begin{equation}
\lb{sema}
\sum_i P_i \frac{\s_i(\o)-\s(\o)}{\s_i(\o)+\s(\o)}.
\end{equation}
Here $\sigma_i(\o)$ denotes the complex conductivity of a local superconducting puddle with stiffness $J_i$ and local $T_c^i$, that are taken with a Gaussian distribution $P_i$ with variance $\sigma_G$ centered around the BCS fit of $J^\mathrm{exp}_s$. For each realization $J_i$ we then compute the $J_{s,i}$ from the solution of the BKT equations \pref{eqk}-\pref{eqg}, we determine the corresponding complex conductivity $\sigma_i(\o)$ and we finally solve Eq.\ \pref{sema} to get the average $J_s^\mathrm{BKT}=(\hbar^2/4e^2)\o\sigma_2(\omega)$ below $T_c$ and the average $\sigma_1(\omega=0)\equiv 1/R^\mathrm{HN}$ above $T_c$, i.e.  the dashed lines reported in Fig.\  4c. Further details about the implementation of the effective-medium approximation can be found in Ref. \cite{ganguly_prb15,venditti_prb19}. The main effect of inhomogeneity is to contribute slightly to the suppression of $J_s$ with respect to $J^\mathrm{BCS}$ before $T_\mathrm{BKT}$. In our case we checked that inhomogeneity, if present, is very small, and a $\sigma_G$ = 0.02 is enough to account for the measured temperature dependences.\\

{\paragraph*{Acknowledgements} {This work was supported by the ANR QUANTOP Project-ANR-19-CE47-0006 grant, by the QuantERA ERA-NET Cofund in Quantum Technologies (Grant Agreement N. 731473) implemented within the European Union's Horizon 2020 Program (QUANTOX) and by Sapienza University of Rome, through the projects Ateneo 2019 (Grant No. RM11916B56802AFE) and Ateneo 2020  (Grant No. RM120172A8CC7CC7), and by the Italian MIUR through the Project No. PRIN 2017Z8TS5B.   } \\

{\paragraph*{Author contributions} N.B. and M.B. proposed and supervised the study. S.M. and H. W. prepared the samples and performed XPS experiments and their analysis. A.G. performed  the STEM and EELS analysis.  G.M., G.S. and S.M. performed the dc and microwave transport experiments and analysed the data with input from M.B, J.L. and N.B..  L.B. conducted the BKT analysis of microwave data.  N.B., M.B and L.B. wrote the manuscript with input from all authors. All authors discussed the results and contributed to their interpretation.} \\


\end{document}